\documentclass[%
 reprint,
 amsmath,amssymb,
 aps,
]{revtex4-2}

\usepackage{graphicx}
\usepackage{dcolumn}
\usepackage{bm}
\usepackage[colorlinks=true, allcolors=black]{hyperref}
\usepackage{comment}
\usepackage{nicefrac}
\usepackage{caption}
\usepackage{subcaption}

\begin{document}

\preprint{APS/123-QED}

\title{Levitodynamics: An analysis of quantum fluctuations based on stochastic mechanics}

\author{Kai-Hendrik Henk}
    \email{kai-hendrik.henk@physik.uni-halle.de}
\author{Wolfgang Paul}%
 
\affiliation{%
Physics Department, Martin-Luther-University Halle-Wittenberg
}%



\date{\today}

\begin{abstract}
Levitodynamics, i.e., the levitation of objects of mesoscopic size has made huge progress in the last decade, giving rise to new experimental opportunities for instance in materials science, but also allowing to address questions of fundamental physics for the first time. It has become possible to cool a levitated particle of mesoscopic size down to its motional ground state and to observe its motion driven by quantum fluctuations. Such an experiment is ideally suited for an analysis within the stochastic mechanics approach to quantum mechanics, which allows for a dynamic description of individual particle paths. We show that within our approach we reproduce the findings of a recent experiment \cite{Magrini_2021}. The phase space dynamics reported there based on a quantum optimal control using Kalman filtering can be understood using the concept of coherent states.

\end{abstract}

\maketitle


\section{Introduction}
Levitation of particles of mesoscopic size (around $100$~nm) by optical, electrical or magnetic interactions has made huge progress in the last decade \cite{Tebben-2021, delic-2020, gonzales-2021, Magrini_2021}. We are especially interested in the possibilities to address foundational questions in quantum mechanics that this progress makes possible. Most notably, it has been reported that a cooling of the center of mass motion of the levitated particle down to its motional ground state can be achieved \cite{Magrini_2021}. An optically trapped silica particle of $70$~nm diameter was cooled close to its ground state by an optimal quantum control implemented through Kalman filtering. The cost function of this experimental optimal control approach was the energy of the quantum system. This work followed the quantum dynamics of one (almost visible) particle in phase space in real time.

Such an experiment is ideally suited to be modeled within the stochastic mechanics approach to quantum mechanics, which contains positions and paths of individual quantum particles as part of its mathematical modeling of quantum systems \cite{nelson1966}. In the next section we will present a short summary of this approach, before mapping the levitation experiment of \cite{Magrini_2021} onto a stochastic mechanics description in section III. In section IV we will present an analysis of the ground state properties of this quantum system and its stationary time-autocorrelation behavior which has been measured experimentally. Section V will then analyze the experimental cooling process in terms of the coherent states of this quantum particle and section VI will present some conclusions.

\section{A short summary of stochastic mechanics}
In 1966, Edward Nelson postulated a description of the motion of quantum particles by a time inversion invariant diffusion process\cite{nelson1966}
\begin{eqnarray}
    dx &= (v(x,t) + u(x,t))dt +dW_{t+}\nonumber\\
    dx &= (v(x,t) - u(x,t))dt +dW_{t-}\, .
    \label{nelson-diffs}
\end{eqnarray}
The first equation is forward in time, the second backward in time with $v(x,t)$ being the drift velocity $u(x,t)$ being the osmotic velocity and $dW_{t\pm}$ being increments of the forward/backward Wiener process. Physically, this models quantum systems as open systems described stochastically, however, the coupling to the environment is not dissipative but conservative (the expectation value of the energy is a constant). This is captured by a stochastic Newton law, $m \overline{a}=F$ where $\overline{a}$ is a suitably adapted definition of acceleration for the continuous but not differentiable paths generated by Eqs.(\ref{nelson-diffs}). Nelson showed that the Schrödinger equation naturally results as the Hamilton-Jacobi description of this Newtonian stochastic dynamics.
In Schrödinger wave mechanics, the entire information on the system is contained in the wave function $\Psi(x,t)$ (we are only needing a one-dimensional description in the following). In the stochastic mechanics formulation of quantum mechanics, this information is also contained in the two velocity fields $v(x,t)$ and $u(x,t)$. 
Using the Madelung representation \cite{madelung} of $\Psi$
\begin{align}
    \Psi(x,t) = \sqrt{\rho(x,t)}e^{\frac{i}{\hbar}S(x,t)}
\end{align}
the velocity fields can be calculated as
\begin{eqnarray}
    u(x,t)&=\frac{\hbar}{2m}\nabla\ln{\rho(x,t)}\label{u_rho}\nonumber\\
    v(x,t) &= \frac{1}{m}\nabla S(x,t).
\end{eqnarray}

In 1995, M. Pavon formulated a quantum Hamilton principle \cite{pavon}
\begin{equation}
  \begin{aligned}
    J[\hat{v},\hat{u}] = \min_v \max_u E \biggr\{ \int \mathrm{d}t \biggr[ \frac{1}{2}m[v(x,t) \\
    -iu(x,t)]^2 - V(x,t) \biggr] + \Phi_0 \biggr\}
  \end{aligned}
\end{equation}
as an extension of the Hamilton principle of classical mechanics. Here $\Phi_0$ formally captures some initial cost value and the Lagrangian has a complex valued kinetic energy term. This complex valued formulation is a compact way to capture two extreme principles which have to be fulfilled at the same time, a stationary action principle as the real part of this quantum Hamilton principle, and a stationary entropy production principle as its imaginary part \cite{pavon}.  Based on this principle, Köppe et al. \cite{koeppe-2017} derived the quantum Hamilton equations 
\begin{eqnarray}
 dx(t) = \left[v(x,t)+u(x(t))\right]dt + \sqrt{\frac{\hbar}{m}} dW_{t+} \\
 dp(x(t)) = \frac{1}{m}V(x(t)) dt + 2\sqrt{\hbar m} \partial_x u(x(t)) dW_{t-}\,.
\end{eqnarray}

For the ground state of a quantum system the drift velocity is zero and the quantum Hamilton principle simplifies to 
\begin{equation}
\begin{aligned}
    J[u_0] = \max_u E\biggr\{ \int_0^T \mathrm{d}t \biggr[-\frac{1}{2}mu^2(x(t)) \\
    - V(x(t))\biggr] +S_0\biggr\}\;.
    \label{gs-opt-cont}
\end{aligned}
\end{equation}
Using stochastic optimal control theory\cite{koeppe-2017, oksendal} one derives the ground state control equations
\begin{equation}
\begin{aligned}
 \mathrm{d}x(t) = u(x(t))\mathrm{d}t + \sqrt{\frac{\hbar}{m}} \mathrm{d}W_{+}(t)  \label{qhe_motion_x_stationary}
\end{aligned}
\end{equation}
for the forward propagation of the position and
\begin{equation}
\begin{aligned}
du(x(t)) = \frac{1}{m}V(x(t)) dt + \sqrt{\frac{\hbar}{m}} \partial_xu(x(t)) dW_{-}(t)
    \label{qhe_motion_u_stationary}
\end{aligned}
\end{equation}
for the backward propagation of the osmotic velocity. The ground state optimal control problem of Eq.(\ref{gs-opt-cont}) is equivalent to the Rayleigh-Ritz variation principle of quantum mechanics \cite{algorithm}, i.e., it is a control aiming to minimize the energy. The optimal control problem in Eq.(\ref{gs-opt-cont}) thus is the mathematical equivalent of the optimal quantum control realized in the experiment \cite{Magrini_2021}. We solve the optimal control problem numerically finding the optimal control osmotic velocity using a genetic algorithm and a parametrization of the osmotic velocity with feed-forward neural networks \cite{algorithm}.

\section{Modeling the levitation experiment}
In the experiment \cite{Magrini_2021} a dielectric silica sphere of diameter $70$~nm with a mass of $m\approx 2.8\, 10^{-18}$~kg is trapped in the intensity maximum of the focus of a laser beam. The experiment reports anisotropic oscillation frequencies in the potential minimum created by the laser particle interaction of $\omega_y/2\pi=305$~kHz, $\omega_z/2\pi=236$~kHz and $\omega_x/2\pi=104$~kHz. The x-direction is the one along the laser beam, which we will focus on in the following. 
\begin{figure}
    \begin{center}
        \includegraphics[width=\columnwidth]{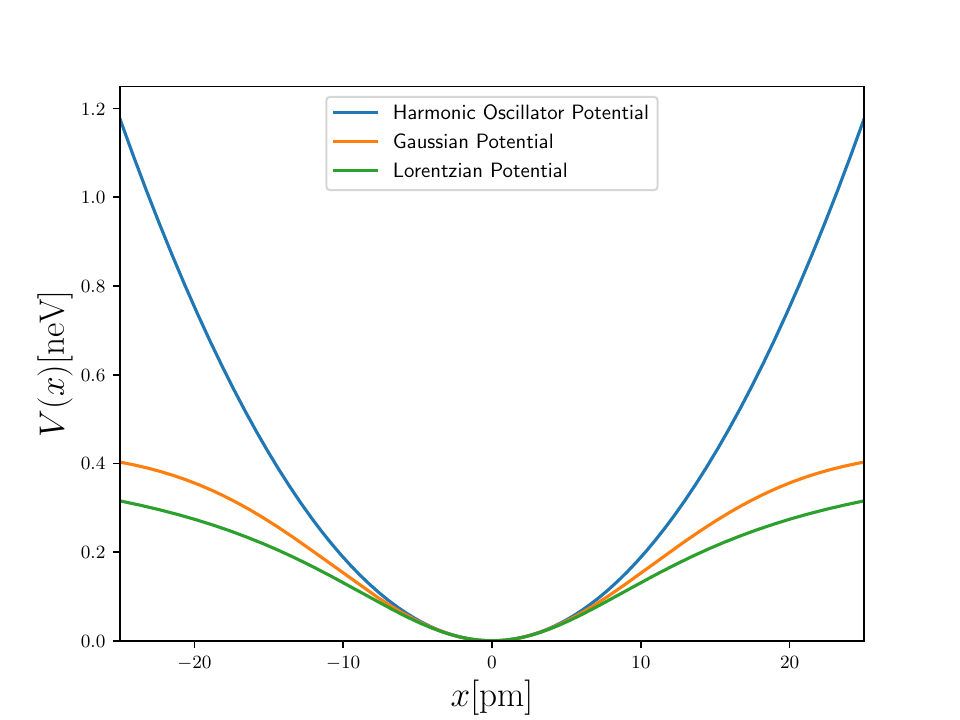}
    \end{center}
    \raggedright
    \caption{\raggedright The three different potentials analyzed in this paper.}
    \label{fig:potentials}
    
\end{figure}The form of the confining potential is given by the shape of the laser beam which is often approximated as Gaussian.
\begin{align}
    V_G(x) &= V_0 \left( 1-e^{-\frac{x^2}{2x_0^2}}\right)
\end{align}
or Lorentzian 
\begin{align}
    V_L(x) &= V_0 \left( 1-\frac{1 }{\frac{x^2}{2x_0^2} + 1}\right)\,.
\end{align}
Both can be approximated by a harmonic potential
\begin{align}
    V(x) = \frac{m\omega^2}{2}x^2=\frac{V_0}{2}\frac{x^2}{x_0^2}\,.
\end{align}
In all three forms $V_0$ sets the energy scale and $x_0$ the length scale. The three potentials are compared in Fig(\ref{fig:potentials})).
One can see, that Gaussian and Lorentzian induce a softer confinement of the fluctuations than their harmonic approximation.
The typical length scale for the harmonic oscillator is $x_0^2= \frac{2\hbar}{m\omega}$ setting the energy scale to $V_0=\hbar\omega$. Values for all scaling variables are given in appendix A.
\begin{figure}
    \centering
    \includegraphics[width=\columnwidth]{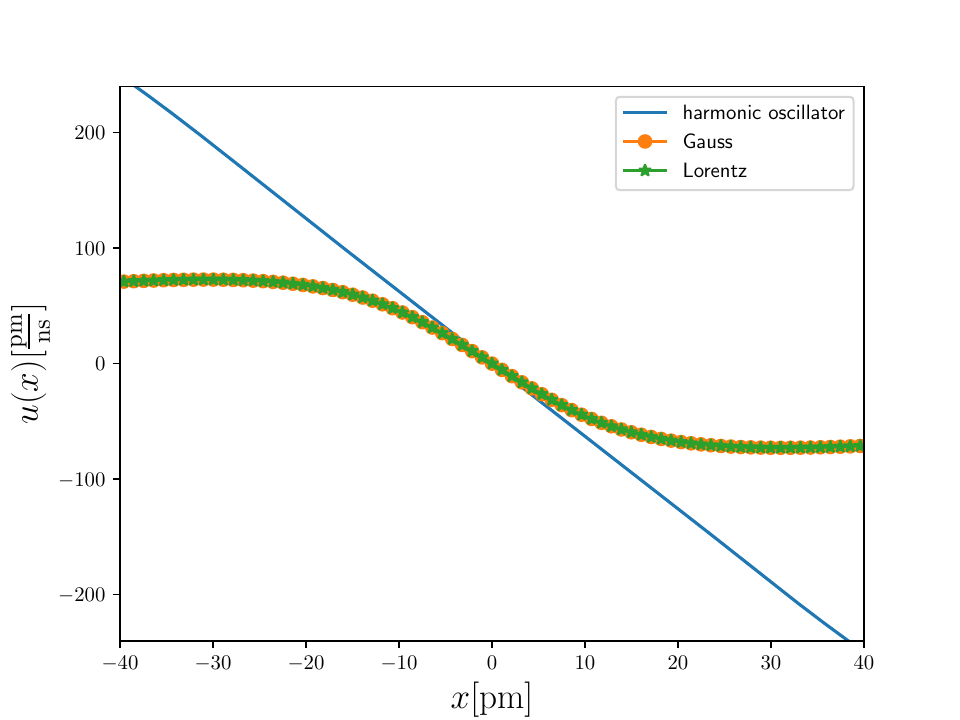}
    \raggedleft
    \caption{\raggedright The three osmotic velocities for the corresponding potentials. The velocities for the Gaussian and the Lorentzian potential are quite similar to one another. The one for the harmonic oscillator is analytically known to be $u(x)=-x$.}
    \label{fig:u_x}
\end{figure}

\section{Ground state properties}
The ground state energies and osmotic velocities can be determined finding the minimum of the stochastic energy
\begin{align}
    \mathbb{E}[E] = \int_{-\infty}^\infty\biggl(\frac{m}{2}u(x)^2 + V(x)\biggr)\rho(x)dx
\end{align}
where the ground state probability density is calculated from the osmotic velocity using 
\begin{equation}
    \rho_0(x) = \exp\left\{\frac{2m}{\hbar}\int^x u_0(x')\mathrm{d}x'\right\}\;.
\end{equation}
The ground state osmotic velocities are shown in Fig.~\ref{fig:u_x}.
The resulting energies can be seen in table \ref{tab:energies}.  
\begin{figure}
\begin{tabular}{c|c|c|c}
\label{energies}
Potential &$E_0$ & $\Delta x \Delta p $ & $ \frac{\Delta x \Delta p}{0.5} $\\
\hline
       harm. Osc. & $0.50004$ & $0.48500$ & $0.97$\\
      Gaussian & $0.40559$ & $0.83675$ & $1.67$\\
       Lorentzian &$0.36484$ &$0.93271$& $1.86$
\end{tabular}
\caption{Ground state energies (in units of $\hbar\omega$) of the three potentials, product of position and momentum uncertainties (in units of $\hbar$) and their ratio to the expected harmonic oscillator value.}
\label{tab:energies}
\end{figure}
As expected, the ground state energies in the Gaussian and the Lorentzian potential are lower than in the harmonic oscillator as these potentials allow for larger quantum fluctuations. Correspondingly, the position-momentum uncertainty in the ground state of these potentials is larger than for the harmonic oscillator, which is a minimum uncertainty state. These uncertainties are determined by simulating sample paths for the ground state in the potentials. In reduced units, the equations of motion in Eqs.(\ref{qhe_motion_x_stationary}) and (\ref{qhe_motion_u_stationary}) read
\begin{eqnarray}
d\tilde{x} &=&  \tilde{p}dt + dW_{+}(t)\\ 
d\tilde{p} &=& \tilde{V}'(\tilde{x}) dt + \tilde{u}'(\tilde{x})dW_{-}(t)\;,\nonumber\label{qhe}
\end{eqnarray}
where the $\tilde{u}'$ are the derivatives of the functions shown in Fig.\ref{fig:u_x}. 

In order to create sample paths for the three potentials, we have to turn the momentum equation into a forward equation. We use Itô's formula to calculate the forward evolution of the momentum (equivalently in scaled variables, the osmotic velocity
\begin{eqnarray}
du(x(t)) &=& u'(x(t)) dx(t) + \frac{1}{2} u''(x(t)) \left(dx(t)\right)^2 \nonumber\\
&=& \left[ u'(x(t))u(x(t)) 
+ \frac{1}{2} u''(x(t))\right] dt \\&&+ u'(x(t)) dW_{+}(t)\nonumber
\end{eqnarray}
One can directly see that in reduced units, for the harmonic oscillator with $u(x)=-x$ position and momentum equations completely decouple
\begin{align}
    du(t) = -u(t)dt -dW_{t+}\label{ho_u}
\end{align}
which is the Ornstein-Uhlenbeck process, the only stationary Gaussian Markov processes. We obtain the same process for the position of the harmonic oscillator
\begin{align}
    dx(t) = -x(t)dt +dW_{t+}.\label{ho_x}
\end{align}
Their respective position and momentum distributions are identical Gaussians (in reduced units), a hallmark of the minimum uncertainty ground state in the harmonic oscillator.

The velocity and position processes of the other potentials do not decouple and we have to solve two coupled forward stochastic differential equations
    \begin{align}
    dx(t)&=u(x(t))dt +dW_{t+}\label{station-forward}\\
    du(t)&=\bigg(\frac{\partial u(x(t))}{\partial x} u(t) +\frac{1}{2}\frac{\partial^2u(x(t))}{\partial x^2}\bigg)dt \\
    &+\frac{\partial u(x(t))}{\partial x}dW_{t+}\nonumber\;.
\end{align}

\begin{figure}
    \centering
    \includegraphics[width=\columnwidth]{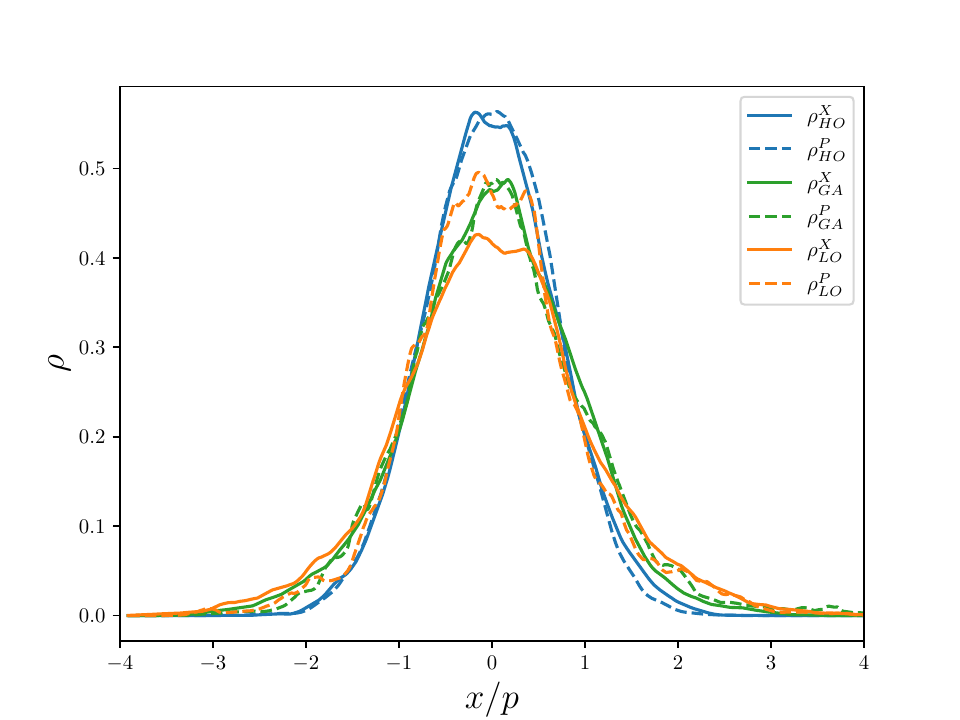}
    \caption{\raggedright Probability distributions for position and momentum. Both overlap well for the harmonic oscillator as they should in reduced units. The distributions of the position for the Gaussian and Lorentzian  potentials and for their momenta also overlap but are broader.}
    \label{fig:densities}
\end{figure}

These two equations are solved numerically with the Heun-scheme~\cite{heun_1}. We integrated Eqs.(\ref{station-forward}) with a time step $\Delta t=10^{-4}$ for $10^7$ steps with $\tilde{x}(0) = 0$ and $\tilde{p}(0)=0$ as initial conditions. These initial conditions  are the expectation values for the position and momentum for all three potentials. The probability densities were obtained as visitation histograms (Fig.~\ref{fig:densities}). 
From the standard deviations of these probability densities, we obtain the position-momentum uncertainties shown in table \ref{tab:energies}.

\subsection{Time Series Analysis}
The solutions of these coupled stochastic differential equations, the sample paths $x(t)$ and $p(t)$ are time series and can thus be analyzed using standard time series analysis methods \cite{brockwell_1, brockwell_2}. 
\begin{figure}
    \centering
    \includegraphics[width=\columnwidth]{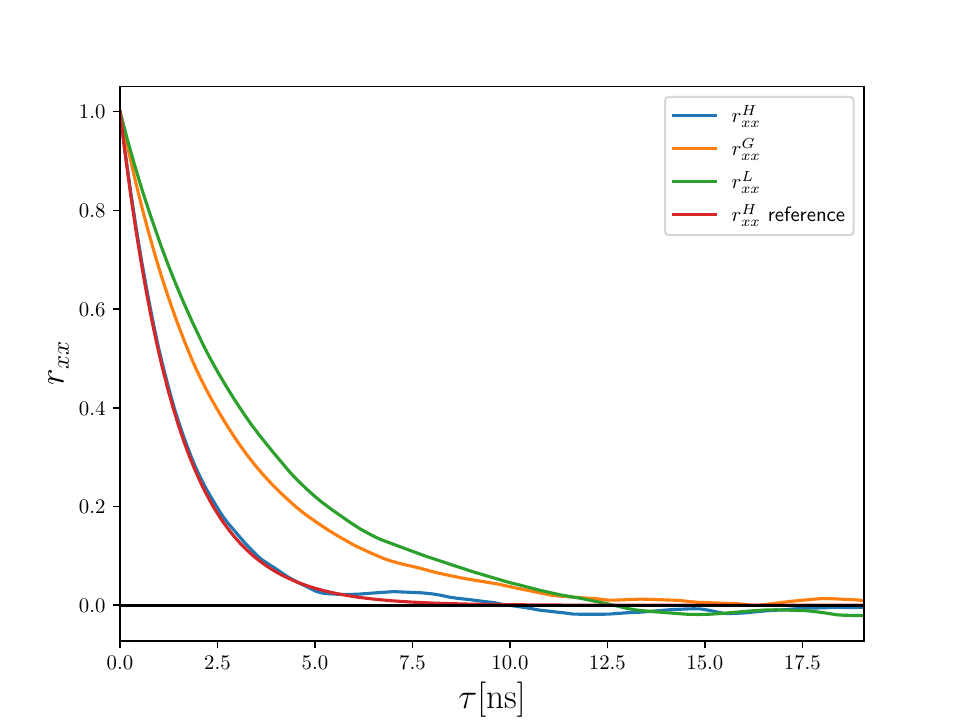}
    \caption{\raggedright Position autocorrelation function for the three potentials. The reference for the harmonic oscillator is the analytic ACF for the Ornstein-Uhlenbeck process.}
    \label{fig:ACF}
\end{figure}
For a stationary stochastic process, its autocorrelation function (ACF) is defined as
\begin{align}
    r_{XX}(\tau) = \frac{\mathbb{E}[(x(t)-\mu)(x({t+\tau})-\mu)]}{\sigma^2}
\end{align}
where $\mu$ is the average value of $x(t)$\cite{brockwell_1, brockwell_2}. The ACF shows how much a signal is correlated with itself over a time difference $\tau$, which does not depend on the reference time for stationary problems. The ACF starts at $r_{XX}(\tau=0)=1$ and converges to $\lim_{\tau \rightarrow \infty}r_{XX}(\tau)=0$ per the above definition. 
Using the Wiener-Khinchin-theorem~\cite{wiener_khinchin}, we can get the power spectral density (PSD) $S_{xx}(f)$ by Fourier-transforming the ACF
\begin{align}
    S_{xx}(f) &= \int_{-\infty}^{\infty} r_{XX}(\tau)\cos(2\pi f\tau)\mathrm{d}\tau.
\end{align}
The PSD gives information about how much the different frequencies add to a signal, which can be used in an experimental setting to find sources of and reduce noise in the signal.
\begin{figure}
    \centering
        \includegraphics[width=\columnwidth]{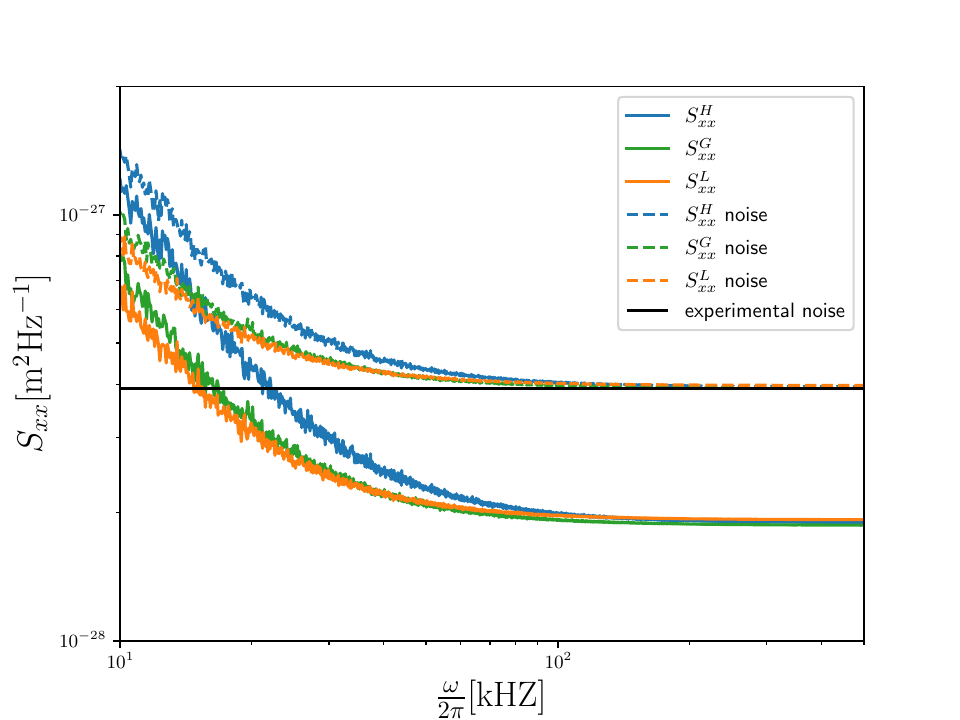}
        \label{fig:PSD}
    \caption{\raggedright Power spectral density of position fluctuations in the three potentials with and without added residual experimental noise \cite{Magrini_2021}.}
    \label{fig:PSD}
\end{figure}
For the determination of the ACF and PSD, we propagated our particles with the discrete time step $\Delta t = 10^{-4}$ for $T=5\cdot10^7$ time steps. Then we calculated the ACF for each window with starting point of every 1000th step and the last value being the end of the time series and averaged over all windows to obtain our results (FIG.\ref{fig:ACF}) and used Fast-Fourier-transformation to get the PSD (FIG.\ref{fig:PSD}). Frequency scale and range of the PSD are in excellent agreement with the experimental results in \cite{Magrini_2021} without any free parameters. Over the first $100$kHz, the signal decays into the residual experimental noise. The harmonic oscillator ACF decorrelates faster than the two other ACFs. The Gauss and the Lorentz ACF don't differ much from each other. The harmonic oscillator PSD decays slower than the other two PSDs, which are quite similar to each other. The small difference from the ACFs also propagates to the PSDs. Considering the magnitude of the residual noise in the experiment\cite{Magrini_2021}, the differences in the PSDs should not be measurable. 

\section{Phase Space Dynamics}
Another fascinating result of the experiments in \cite{Magrini_2021} is their ability to reconstruct phase-space trajectories of the levitated particle from the quantum optimal control approach. The authors stress that the resolution of these trajectories does not violate Heisenberg's uncertainty relation.

\begin{figure}
        \centering
            \includegraphics[width=\columnwidth]{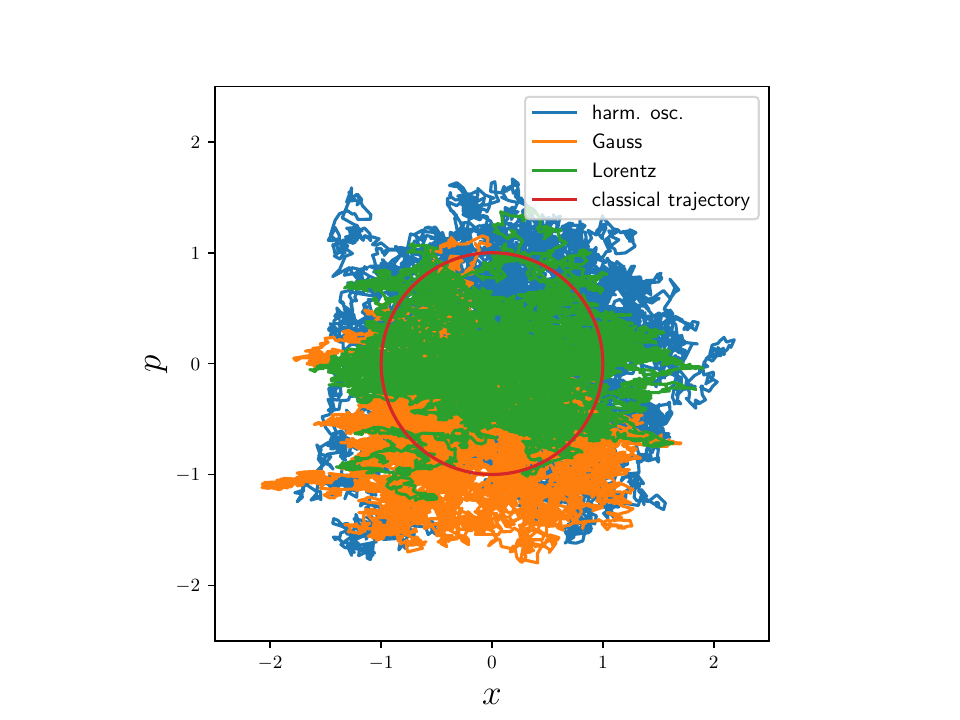}

    \caption{\raggedright Phase space sample paths for ground state processes in the three potentials. The classical trajectory is given for the ground state energy $\nicefrac{1}{2} \hbar\omega$.
    }
    \label{fig:phase}
\end{figure}

Using the equations (\ref{station-forward}), we can propagate $x(t)$ and $p(t)$ separately forward in time. Thus we can combine them to get sample paths in phase space as shown in Fig.\ref{fig:phase}. 

The phase-space trajectories for the ground state processes in the three potentials are shown in Fig.~\ref{fig:phase}. 
The ground state processes densely cover a roughly circular phase-space area. Qualitatively, one sees the same in the experiment \cite{Magrini_2021} for the trajectory of the system close to its motional ground state. 
In \cite{Magrini_2021}, the trajectory of the system during the cooling process is also shown. We will describe this cooling process based on the coherent states of the harmonic oscillator.

\subsection{Coherent States}
\begin{figure}
   
   \begin{center}

    \begin{subfigure}{0.4\textwidth}
    \centering
        \includegraphics[width=\columnwidth]{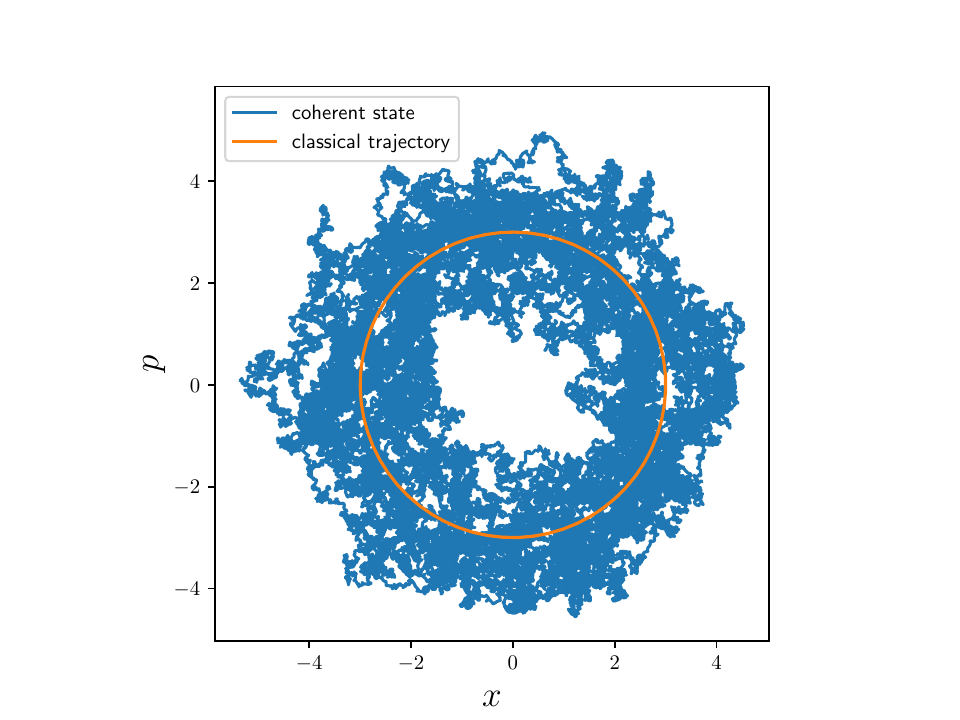}
        \caption{}
        \label{fig:coherent}
    \end{subfigure}
    \begin{subfigure}{0.4\textwidth}
    \centering
        \includegraphics[width=\columnwidth]{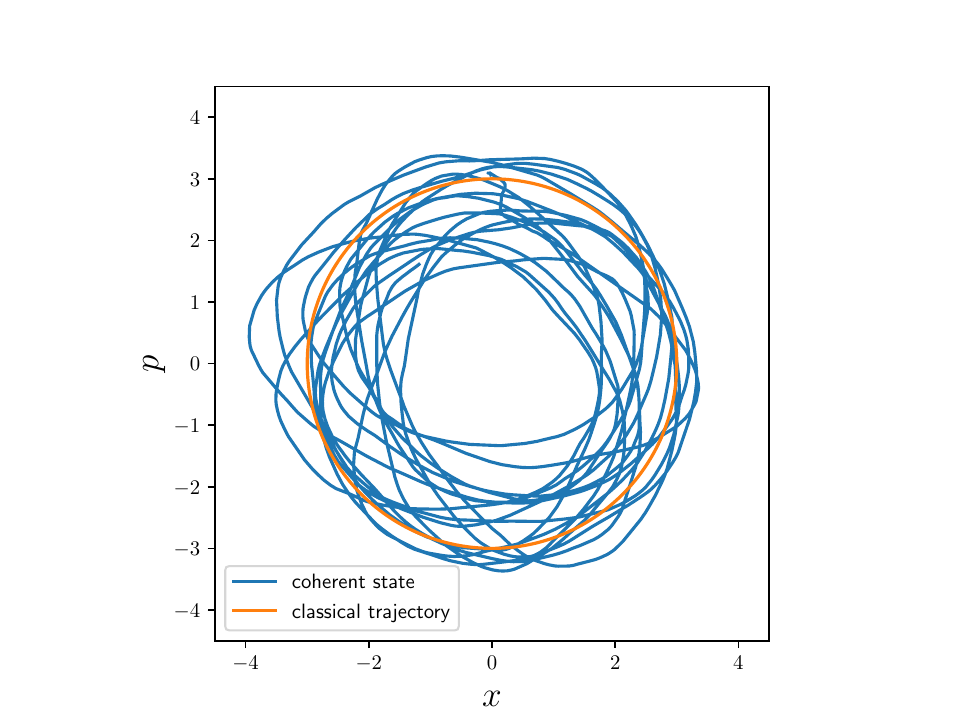}
        
        \caption{}
        \label{fig:coherent_tma}
    \end{subfigure}
    \end{center}
    \caption{\raggedright Phase space trajectory of a coherent state with classical energy $E=\nicefrac{3}{2}\hbar\omega$ without (a) and with (b) a time-moving average for smoothing applied. The averaging is done over $10^4$ points to reflect the $32$~ns time resolution of the experiment\cite{Magrini_2021}.} 
\end{figure}

Coherent states, sometimes called Glauber states, first proposed by Schrödinger in 1926~\cite{schrödinger_coherent}, are bound, non-stationary states, that in the classical limit behave like the equivalent classical state of the harmonic oscillator~\cite{glauber_1, klauder}. In \cite{paul2013stochastic} the stochastic mechanics realization of the coherent states is discussed. The wave function in a coherent state contains the classical phase space trajectories $x_{cl}(t), p_{cl}(t)$ 
\begin{align}
    \psi (x,t) = (2\pi \tilde{\sigma})^{-\frac{1}{4}}e^{-\frac{(x-x_{cl}(t))^2}{4\tilde{\sigma}^2} + \frac{i}{\hbar} xp_{cl}(t) - \frac{i}{2\hbar}x_{xl}(t)p_{cl}(t)-\frac{i}{2}\omega t}
\end{align}
with $\tilde{\sigma} = \frac{\hbar}{2m\omega}$. The drift and osmotic velocities therefore are
\begin{align}
    v(x,t)&=\frac{p_{cl}(t)}{m}\\
    u(x,t)&=-\omega(x-x_{cl}(t))
\end{align}
resulting for the propagation in space in
\begin{align}
    dx=\biggl(\frac{p_{cl}(t)}{m} -\omega(x-x_{cl}(t)) \biggr)dt +\sigma dW_+.\label{coherent_x}
\end{align}
The coherent states are also solutions to the quantum Hamilton equations for the harmonic oscillator Eqs.(\ref{qhe}). However, we will again use a forward equation for the momentum
\begin{align}
    p=m(v(x(t),t)+u(x(t),t))\;.
\end{align}
From the Itô-Formula we get
\begin{equation}
\begin{aligned}
    dp(t) &= (\frac{\partial p}{\partial x}\biggl(\frac{p_{cl}(t)}{m} -\omega(x-x_{cl}(t))\biggr) \\&+M\biggl( \frac{\partial x_{cl}(t)}{\partial t} +\frac{1}{M}\frac{\partial p_{cl}(t)}{\partial t}\biggr) )dt +\frac{\partial p}{\partial x}dW_+
\end{aligned}
 \end{equation}
\begin{align}
    &=(-\omega p(t) +M\biggl( \frac{\partial x_{cl}(t)}{\partial t} +\frac{1}{M}\frac{\partial p_{cl}(t)}{\partial t}\biggr))dt -\omega dW_+.\label{p_coherent_ito}
\end{align}
For the simulations shown in Fig.\ref{fig:coherent}, we use a classical trajectory with starting conditions  $x_{cl}(0) = 0, p_{cl}(0) = 3$, i.e. an energy of the classical motion of $3\hbar\omega$. The coherent state motion consists of the classical trajectory
on top of which one has the quantum fluctuations of the ground state, resulting in minimum uncertainty fluctuations around the classical trajectory. 
 In the experiment \cite{Magrini_2021}, the Kalman-Filter electronics was able to determine position and momentum every $32$~ns, so we use a time-moving average over this acquisition time as well (see Fig.~\ref{fig:coherent_tma}). 

\subsection{Modeling the Cooling Process}
In \cite{Magrini_2021}, the phase space trajectory along the cooling process is also shown. To model this, we replaced the conservative, classical trajectory of the coherent state by that of a damped harmonic oscillator
\begin{align}
    x_{cool} (t) &= x_{cl}(t)\cdot e^{-t/t_d} = 3\sin (\omega t)\cdot e^{-t/t_d} \\
    p_{cool} (t) &= p_{cl}(t)\cdot e^{-t/t_d} = 3\cos (\omega t)\cdot e^{-t/t_d} ,
\end{align}
with the damping time equal to the reaction time of the Kalman filter, $t_d=32$~ns. The shift of the oscillation frequency can be neglected for this choice of damping.
\begin{figure}
    \begin{center}
    \includegraphics[width=\columnwidth]{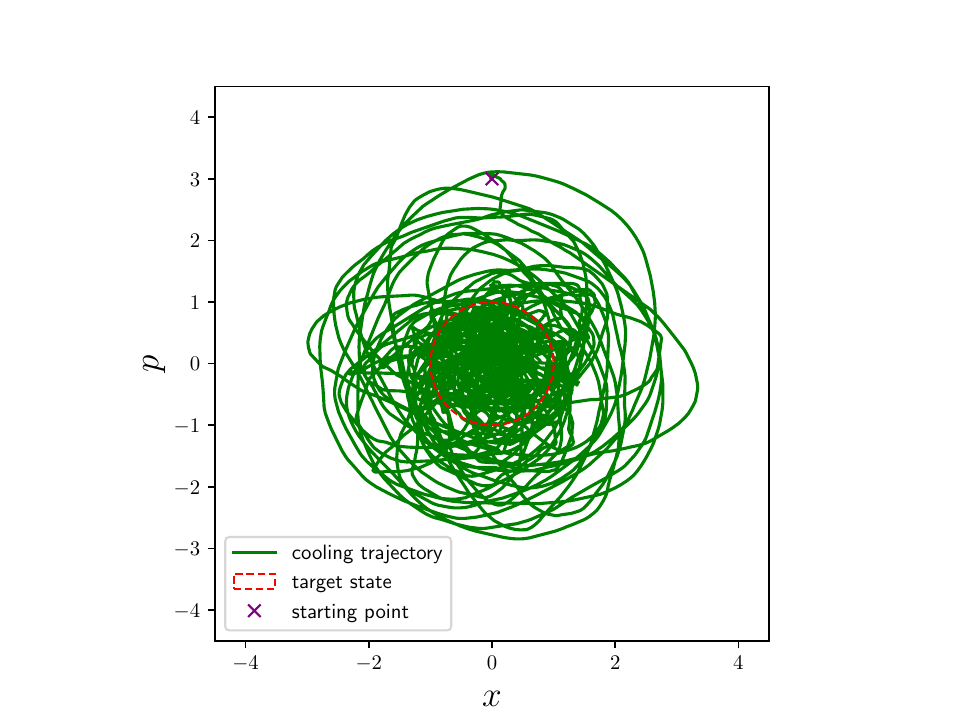}
    \end{center}
    \caption{\raggedright Phase space trajectory of the cooling process with the time moving average already applied.}
    \label{fig:cooling}
\end{figure}
The result can be seen in Fig.\ref{fig:cooling}. The cooling trajectory closely resembles the cooling trajectory for energy $E=2.71\hbar\omega$ shown in \cite{Magrini_2021}. However, we cool down completely to the quantum mechanical ground state. 

\section{Conclusion}
In this paper, we have shown how by using stochastic mechanics one can 
model the dynamics of levitated particles of mesoscopic size in their quantum ground state. Using the information given in \cite{Magrini_2021} it was possible to quantitatively reproduce time series statistics like the power spectral density which had been measured experimentally.
Additionally we have shown, that it is not only possible but also useful to draw phase portraits of quantum systems without violating the Heisenberg uncertainty principle. We have also discussed, that for experimental purposes, the main difference between the actual form of the tweezer potentials and a harmonic approximation thereof are the ground state energies and the values of the position-momentum uncertainties in the ground state.

\appendix


\section{Units\label{units}}
For simulation purposes dimensionless units were used. For space, momentum and energy we use the standard reference values of the harmonic oscillator $x_0=\sqrt{\frac{\hbar}{m \omega}},\; p_0=\sqrt{\hbar m \omega},\;E_0=\hbar\omega$. The tweezer potentials depend on the intensity of the laser light but we know their strength from the induced oscillation frequency reported in \cite{Magrini_2021}
\begin{align}
    m &\approx 2.8\cdot10^{-18}\\
    \omega &=2\pi\cdot10^4\mathrm{kHz}
\end{align}
With that we can report our results in SI-units using
\begin{align}
    t_0 &= 0.159\cdot10^{-9}s\\
    x_0 & = 7.742\cdot10^{-13}m\\
    p_0& =1.884\cdot10^{-19}\mathrm{kg}\frac{\mathrm{m} }{\mathrm{s}}\\
    u_0&=\frac{p_0}{m}=67.3\frac{\mathrm{pm} }{\mathrm{ms}}
\end{align}
For the analysis of the phase-space dynamics, we still used reduced units.

\section{Hard- and Software Implementation}
The programs were run on a Intel Core i7-10700 CPU @ 2.90GHz × 16 on a Ubuntu 20.04.5 LTS 64 bit system using PyCharm Community Edition. For the discretized Wiener process, the random number generator PCG64 \cite{pcg} with seed $2168461$ from the numpy package was used. The programs were programmed in Phyton 3.8 with the numpy and matplotlib packages.


%

\end{document}